\documentclass[
aps,%
final,%
notitlepage,%
oneside,%
twocolumn,
nobibnotes,%
nofootinbib,
superscriptaddress,%
noshowpacs,%
centertags]%
{revtex4}
\usepackage[T2A]{fontenc}
\usepackage[english]{babel}
\usepackage{amsfonts}\usepackage{amsbsy}
\usepackage{color}
\usepackage{amsfonts}
\usepackage{amsbsy}
\usepackage{mathrsfs}
\usepackage{graphicx}
\def\lsim{\mathrel{\rlap{
\lower4pt\hbox{\hskip-3pt$\sim$}}
    \raise1pt\hbox{$<$}}}     
\def\gsim{\mathrel{\rlap{
\lower4pt\hbox{\hskip-3pt$\sim$}}
    \raise1pt\hbox{$>$}}}     
\def\scr#1{\mbox{\scriptsize #1}}
\begin{document}
\title{
On freeze-out problem in relativistic
  hydrodynamics\footnote{Dedicated to 
  S.T. Belyaev on the occasion of his 85th birthday.}}%
\author{\firstname{Yu.B.~Ivanov}}
\email[]{Y.Ivanov@gsi.de}
\altaffiliation{RRC ``Kurchatov Institute'', Kurchatov sq.$\!$ 1, Moscow
123182, Russia}
\affiliation{Gesellschaft f\"ur Schwerionenforschung mbH, Planckstr.$\!$ 1,
64291 Darmstadt, Germany}
\author{\firstname{V.N.~Russkikh}}
\email[]{russ@ru.net}
\altaffiliation{RRC ``Kurchatov Institute'', Kurchatov sq.$\!$ 1, Moscow
123182, Russia}
\affiliation{Gesellschaft f\"ur Schwerionenforschung mbH, Planckstr.$\!$ 1,
64291 Darmstadt, Germany}
\begin{abstract}
A finite unbound system which is equilibrium in one reference frame is in
general nonequilibrium in another frame. This is a consequence of the
relative character of the time synchronization in the relativistic
physics. This puzzle was a prime motivation 
of the Cooper--Frye  approach to the freeze-out in 
relativistic hydrodynamics. 
Solution of the puzzle reveals that the
Cooper--Frye recipe is far not a unique phenomenological method that
meets requirements of energy-momentum conservation. Alternative freeze-out
recipes are considered and discussed. 
%
\end{abstract}
%
\maketitle

\section{Introduction}

Hydrodynamics is now a conventional approach to simulations
of heavy-ion collisions. Even review papers
\cite{Clare,Stoecker86,MRS91,Rischke98,Kolb04,Ruuskanen06} do not 
comprise a complete list of numerous applications of this approach.
The hydrodynamics is applicable to description of hot and dense
stage of nuclear matter, when the mean free path is well shorter
than the size of the system. However, as expansion proceeds, the system
gets dilute, the mean free path becomes comparable to
the system size, and hence the hydrodynamic calculation should be
stopped at some instant. All hydrodynamic calculations are
terminated by a freeze-out procedure, while these freeze-out
prescriptions are somewhat different in different models.
Moreover, the freeze-out prescriptions include recipes to calculate 
spectra of produced particles which are of prime experimental 
interest.

Historically the first method for freeze-out was suggested by Milekhin 
\cite{Milekhin} in the context of the Landau hydrodynamic model of 
multiple production of particles in high-energy hadron collisions 
\cite{Landau53}. Later, Milekhin's approach was criticized by 
Cooper and Frye \cite{Cooper}. Cooper and Frye pointed out that  
Milekhin's approach does not conserve energy and proposed their own 
recipe of the freeze-out. In this paper 
we would like to discuss a puzzle which was in fact a prime motivation 
of the Cooper--Frye approach \cite{Cooper} to the freeze-out in the 
relativistic hydrodynamics. This puzzle is closely related to the
definition of the relativistically invariant distribution function 
as it was for the first time advanced by S.T. Belyaev and G.I. Budker
\cite{BB56}.

\section{The puzzle}

Let us consider a droplet of matter (for simplicity 
consisting of only nucleons),  
which is 
characterized by a total baryon number $N$, 
a total energy $E$ and a total momentum ${\bf P}$, and occupies a volume
$V$. To be precise, we assume that this droplet is a closed system.

Let this droplet be 
described by an equilibrium distribution (in configuration and
momentum space)
\begin{eqnarray}
\label{f-eq} 
f (x,p) =
\frac{g}{(2\pi)^3}%
\frac{1}{\exp\left\{\left(p_\mu u^\mu - \mu \right)/T\right\}+ 1}
\end{eqnarray}
in the reference frame charaterized by 4-velocity $u^\mu$. Let us call
this frame as a computation one\footnote{ i.e. that where the
hydrodynamic computation takes place. 
}. 
This distribution is 
defined in terms of degeneracy of the nucleon $g$, 
chemical potential $\mu$, temperature $T$ and already mentioned
4-velocity $u^\mu$.  
The 4-velocity $u^\mu$ is commonly interpreted as a velocity with
which the droplet moves as a whole. 
We asssume 
that this distribution is 
homogeneous in the volume $V$. The last requirement is an 
important condition of the equilibrium. 
Therefore, the $x$ dependence is in fact absent in Eq. (\ref{f-eq}). 
{\em In particular, distribution function (\ref{f-eq}) defines the way
  how it changes under the Lorentz transformation.}

In terms of this distribution function, the conserved quantities of
the droplet can be expressed as follows. First we calculate baryon
density ($\rho$) 
and elements of the energy-momentum tensor ($T^{\mu\nu}$) in the
computation frame 
\begin{eqnarray}
\label{J0} 
\rho &=& \int \frac{d^3 p}{p^0} p^0 \ f (p), 
\\
\label{T00} 
T^{\mu\nu} &=& \int \frac{d^3 p}{p^0} p^\mu p^\nu \ f (p) 
= (\varepsilon + P) u^\mu u^\nu - g^{\mu\nu} P, 
\end{eqnarray}
where $\varepsilon$ and $P$ are the proper energy density and pressure,
respectively. 
Then we multiply these quantities by the volume $V$
and thus obtain 
\begin{eqnarray}
\label{N} 
N &=& \rho V, 
\\
\label{E} 
E &=& T^{00} V = [(\varepsilon + P) u^0 u^0 - P] V, 
\\
\label{Pi} 
P^i &=&  T^{0i} V = (\varepsilon + P) u^0 u^i V.    
\end{eqnarray}

Now we are able to formulate the puzzle. We know that $(E, {\bf P})$
is a 4-vector, at least this is stated in all textbooks. 
To be precise, the fact that 
\begin{eqnarray}
\label{P4} 
P^\mu = \int dV \ T^{0\mu}
\end{eqnarray}
is indeed a 4-vector and that $P^\mu$ is independent of the
frame\footnote{Experts in the freeze-out prefer to call it as
  independence of the 3D hyposurface in the Minkowski space.}
(up to a Lorentz transformation), where it is calculated, is proved, e.g., 
in Ref. \cite{Weinberg}\footnote{See also Ref. \cite{3FD-FO}, where
  this proof is accommodated to the problem of freeze-out in nuclear
  collisions.}. 
Then the relation 
\begin{eqnarray}
\label{P2E} 
P^i/E \stackrel{?}{=}  u^i/u^0 \equiv v^i
\end{eqnarray}
should take place, if $v^i$ is the velocity of motion of this droplet
as a whole. As we see from Eqs. (\ref{E}) and (\ref{Pi}), this is not
the case. 
{\em 
Than the questions arise: what is the meaning of the
4-velocity $u^\mu$ and
what is the meaning of     
the proper energy density $\varepsilon$ and the pressure $P$?}

Moreover, if we believe that the 4-velocity $u^\mu$ is the velocity of
motion of this droplet as a whole and  $\varepsilon$ is the 
energy density in the droplet-rest frame, we can first calculate 
$P^\mu$ in the droplet-rest  frame (where $u^\mu=(1,0,0,0)$) and
then boost it into the 
computation frame. Then we arrive at another surprising result 
\begin{eqnarray}
\label{E1} 
E &\stackrel{?}{=}& \varepsilon u^0 V^*, 
\\
\label{Pi1} 
P^i &\stackrel{?}{=}& \varepsilon u^i V^*,     
\end{eqnarray}
where $V^*$ is the volume in the droplet-rest frame. 
Now the above puzzle reads as follows. 
There exists no $V^*$ which
makes Eqs. (\ref{E1}) and (\ref{Pi1}) compatible with Eqs. (\ref{E}) and
(\ref{Pi}). 
This again makes us doubtful about interpretation of $u^\mu$,
$\varepsilon$ and $P$ quantities. 

In fact, precisely the contradiction between
Eqs. (\ref{E1})--(\ref{Pi1}), on the one hand, and
Eqs. (\ref{E})--(\ref{Pi}), on the other hand, motivated Cooper and
Frye \cite{Cooper} to suggest their recipe for the freeze-out, which
just avoids this contradiction rather than resolves it.

\section{Resolution of the puzzle}
\label{Resolution}

Let us consider what really happens to the equilibrium distribution 
(\ref{f-eq}) under the Lorentz transformation. It is convenient to
represent this distribution by an ensemble of particles as 
follows\footnote{Such a representation is extensively used in
  Ref. \cite{Weinberg}.}
\begin{eqnarray}
\label{f-en} 
f (x,p) =
\sum_i \delta^3 ({\bf p - p}_i(t)) \ \delta^3 ({\bf x - x}_i(t)), 
\end{eqnarray}
where ${\bf x}_i$ and ${\bf p}_i$ are the coordinate and momentum of
the $i$th particle, respectively. The ${\bf x}_i$ coordinates
homogeneously populate the volume, $V$, of the droplet in the
computation reference frame. 
{\em By definition of the
distribution function, all these particles are considered at the same
time instant $t$.} Integration of this distribution
function over $d^3p\;d^3x$ with weights $1$, $p^0$ and ${\bf p}$ 
gives 
\begin{eqnarray}
\label{EP2} 
N= \sum_i 1, \quad  
E = \sum_i p^0_i, \quad  {\bf P} = \sum_i {\bf p}_i,  
\end{eqnarray}
respectively, which explicitly demonstrates that $(E, {\bf P})$ is
indeed a 4-vector.

Let us transform distribution from the computation frame (\ref{f-eq}),
where it is simulated  
by Eq. (\ref{f-en}),  to the rest frame of the droplet. To do this, we
boost these particles with some    
velocity $-{\bf v}^*$ which certainly differs from $v^i=u^i/u^0$ in
view of consideration   of the previous sect. 
Applying a Lorentz transformation to the ensemble of particles
(\ref{f-en}),  we arrive at 
\begin{eqnarray}
\label{Lt-en} 
\sum_i \delta^3 ({\bf p^* - p^*}_i(t^*_i)) \ \delta^3 ({\bf x^* - x^*}_i(t^*_i)), 
\end{eqnarray}
where quantities marked by $^*$ correspond to the rest frame of the
droplet and are obtaned by the Lorentz
transformation\footnote{For definiteness, we assume that ${\bf v}^*$ is
  directed along the  $x$ axis.}
\begin{eqnarray}
\label{Lorentz} 
t^*&=&x\sinh\psi + t\cosh\psi,  
\cr 
x^*&=&x\cosh\psi + t\sinh\psi,
\quad  y^*=y, \quad  z^*=z,  
\end{eqnarray}
with $\tanh\psi=v^*$. 

We do not call   
 sum (\ref{Lt-en}) a distribution function, since all particles are taken at
different time  
instants $t^*_i$. This is a direct consequence of the Lorentz
transformation---events  
which are simultaneous in one reference frame are not necessarily
simultaneous in  another one.

In order to obtain a distribution function from ensemble
(\ref{Lt-en}), we should  reduce this ensemble to a common time, e.g., 
\begin{eqnarray}
\label{ct} 
t^* = \sum_i t^*_i / N, 
\end{eqnarray}
where $N$ is the number of particles in this ensemble, cf. Eq. (\ref{EP2}). 
To do this, we should move particles forward or backward in time, depending on 
the sign of $t^* - t^*_i$. After this reduction the ensemble  (\ref{Lt-en})
already simulates a distribution function in the droplet-rest frame: 
\begin{eqnarray}
\label{Lt-f-en} 
f (x^*,p^*) =
\sum_i \delta^3 ({\bf p^* - p^*}_i(t^*)) \ \delta^3 ({\bf x^* - x^*}_i(t^*)). 
\end{eqnarray}
Doing this in general case, we have to take 
into account that particles at time $t^*$ in the droplet-rest frame 
have exercised additional (or, vise versa, have not exercised 
all those) interactions as compared to those in the computation frame at
time $t$.  
We will avoid these extra complications assuming that particle do not
interact\footnote{Moreover, if a system is in a bound state, e.g. a cold
  nucleus, these additional/missed interactions restore equilibrium in
  any reference frame. Therefore, in this paper we consider an
  inherently unbound  state of the system.}.  
This case is relevant to the problem of freeze-out. In this case 
\begin{eqnarray}
\label{Lt-p} 
{\bf p^*}_i(t^*) &=&  {\bf p^*}_i(t^*_i), 
\\
\label{Lt-x} 
{\bf x^*}_i(t^*) &=& {\bf x^*}_i(t^*_i) 
+  [{\bf p^*}_i(t^*)/p^*_0(t^*)] (t^* - t^*_i), 
\end{eqnarray}
i.e. the momentum ${\bf p^*}_i$ remains the same, but the coordinate
${\bf x^*}_i$ changes. 

Now we are able to analyze the result of the above
transformation. Let, for the sake of definiteness,  the volume
$V$ be a Lorentz contracted spherical volume [contracted with  
gamma factor $\gamma^*=(1-{\bf v}^{*2})^{-1/2}$].  The 
coordinates ${\bf x}_i(t)$ homogeneously populate this volume. 
Since the ${\bf x}_i(t)$ ensemble is taken at the same time instant,
transformed coordinates   
${\bf x^*}_i(t^*_i)$ homogeneously populate the same but Lorentz
``uncontracted'' volume,  $V^*=\gamma^* V$. Indeed, the linear
transformation (\ref{Lorentz}) preserves the spatial homogeneity of
this ensemble.

When we reduce these coordinates to a common time $t^*$, see Eq. (\ref{Lt-x}), 
some high-momentum particles (in the droplet-rest frame) leave the
$V^*$ volume,  while the most part of low-momentum particles  
remains in this volume. {\em Therefore, the Lorentz transformed
  distribution becomes spatially  
inhomogeneous and thus even nonequilibrium. 
This is purely relativistic effect, associted with relative character
of the time synchronization in the relativistic physics. } 
This effect is closely related to the fact that, if even an unbound
system was equilibrium at the initial time instant, it becomes
nonequilibrium at the next time instant because of inhomogeneous
expansion of the system. 
In particular, this is the reason why we failed to find a volume $V^*$ which
makes Eqs. (\ref{E1}) and (\ref{Pi1}) compatible with Eqs. (\ref{E}) and
(\ref{Pi}). There exists simply no common volume  $V^*$ for all
particles in the droplet-rest frame, if it is assumed to be
homogeneous in the computation frame.

{\em Nevertheless, the conventional interpretation of quantities entering
the equilibrium distribution (\ref{f-eq}) and the way of Lorentz
transformation prescribed by it are valid, if a considered droplet is
an open system surrounded by equilibrium medium. }
Let us transform the distribution (\ref{f-en})
in the computation frame by boosting it with the velocity 
$-{\bf v}=-{\bf u}/u^0$. Now let  the volume
$V$ be a Lorentz contracted spherical volume [contracted with  
gamma factor $\gamma=(1-{\bf v}^{2})^{-1/2}$]. Then 
transformed coordinates   
$\widetilde{\bf x}_i(\widetilde{t}_i)$ homogeneously populate a 
spherical volume,  $\widetilde{V}=\gamma V$. However, in view of
discussion in the previous sect., the total 3-momentum
of the droplet in this ``tilded'' frame is still nonzero, 
$\widetilde{\bf P} \neq 0$. When we reduce these coordinates to a common time
$\widetilde{t}$, similarly to Eq. (\ref{Lt-x}), some particles leave
the $\widetilde{V}$ volume,  
but at the same time other particles come to this volume from the
surrounding medium.  After this ``particle exchange with the medium'' 
the total 3-momentum of the droplet, with already changed particle
content, becomes really zero, and its momentum distribution is really
described by Eq.  (\ref{f-eq}) with $u^\mu=(1,0,0,0)$. 


\section{Practical consequences for the freeze-out}
\label{Practical}

Let us address the question of observable spectrum of particles
originating from the frozen out droplet of matter.  Recollect that
this droplet is 
characterized by the total baryon number $N$, 
and total energy $E$, momentum ${\bf P}$, and volume $V$ 
in the computation reference frame. 
All these quantities are known from solution the hydrodynamic equations. 
Note that thermodynamic quantities, i.e. temperature and baryonic
chemical potential are not directly known from hydrodynamics.

From the above
discussion we see that first we should decide in which reference frame
this droplet is equilibrium. There are many possibilities to do this
choice. 

\subsection{Freeze-out in Computation Frame}
\label{Computation}

The first natural choice is that the droplet is equilibrium in the
computation reference frame. Then we determine the 
chemical potential $\mu$, temperature $T$, 4-velocity $u^\mu$, and
volume  $V$  from Eqs. (\ref{N})--(\ref{Pi}) and an equation of
state (EoS). With all parameters of 
the distribution function (\ref{f-eq}) being defined, the invariant
spectrum of observable particles reads as follows 
\begin{eqnarray}
\label{p0-FO}
\left(E \frac{d N}{d^3 p}\right)_{\scr{comp. frame}} =
V \ p^0 \ f (p,x). 
\end{eqnarray}
This spectrum obeys conservations of the baryon number $N$, 
total energy $E$ and momentum ${\bf P}$. Note that this recipe of the
freeze-out differs  both from the Cooper--Frye one \cite{Cooper} and 
from Milekhin's one \cite{Milekhin}. 

A shortcoming of this recipe is that it is closely related to the 
reference frame of computation. In principle, we could do computation
in a different reference frame. 
Note that an effective freeze-out in kinetic
simulations of heavy-ion collisions occurs in the same manner,
i.e. the history of particle 
collisions is followed in the reference frame of computation. 

\subsection{Freeze-out in Local-Rest Frame}
\label{Local-Rest}

Another natural construction is as follows. Let us start as in
the previous sect., i.e. transform distribution from the computation
frame (\ref{f-eq}), where it is simulated  
by Eq. (\ref{f-en}),  to the droplet-rest frame. To do this, we boost the
system to the      
velocity $-{\bf v}^*=-{\bf P}/E$ which certainly differs from $v^i=u^i/u^0$ in
view of the previous consideration. 
Applying a Lorentz transformation to the ensemble of particles
described by Eq. (\ref{f-en}),  we arrive at ensemble of particles
described by Eq. (\ref{Lt-en}). 
This ensemble still does not simulate 
a distribution function, since all particles are taken at
different time  instants $t^*_i$. 

Since we consider freeze-out process, we are not interested in time
instants of these frozen-out particles. Therefore, we artificially 
attribute the same time instant [say, that of Eq. (\ref{ct})] to all
particles without changing their momenta and coordinates. Then we
arrive at an equilibrium distribution function (\ref{Lt-f-en})
but with 
\begin{eqnarray}
\label{Lt-p-f} 
{\bf p^*}_i(t^*) &=&  {\bf p^*}_i(t^*_i), 
\\
\label{Lt-x-f} 
{\bf x^*}_i(t^*) &=& {\bf x^*}_i(t^*_i) , 
\end{eqnarray}
which differ  from (\ref{Lt-p})--(\ref{Lt-x})  only in definition of 
${\bf   x^*}_i(t^*)$. This distribution takes place in an 
``uncontracted'' volume  $V^*=\gamma^* V$.  

From the practical point of view, we should solve equations 
\begin{eqnarray}
\label{Nm1} 
N &=& \rho^* V^*, 
\\
\label{Vm1} 
V &=& \gamma^* V^*, 
\\
\label{Em1} 
E &=& \varepsilon^* u^{*0} V^*, 
\\
\label{Pim1} 
P^i &=& \varepsilon^* u^{*i} V^*,    
\end{eqnarray}
supplemented by a EoS,  
in order to determine $\mu^*$, temperature $T^*$, 4-velocity $u^*_\mu$, and
volume  $V^*$  
in terms of which the invariant
spectrum of observable particles reads as follows 
\begin{eqnarray}
\label{m-FO}
\left(E \frac{d N}{d^3 p}\right)_{\scr{Milekhin}} =
V^* \ (p_\mu u^{*\mu}) \ f^* (x,p),  
\end{eqnarray}
where $f^* (x,p)$ is the equilibrium distribution function defined in
terms of thermodynamic quantities with superscript $^*$,
cf. Eq. (\ref{f-eq}). 
This spectrum obeys conservations of the baryon number $N$, 
total energy $E$ and momentum ${\bf P}$. 
This method can be called a modified Milekhin's freeze-out, since
equations of the original Milekhin's method (\ref{N})--(\ref{Pi})
certainly differ from (\ref{Nm1})--(\ref{Pim1}). Precisely this method
is used in the model of three-fluid dynamics \cite{3FD-FO,3FD}.

An advantage of this recipe is that the choice of the reference frame is
unique and independent of the frame of computation. 
However, the entropy is not spectacularly conserved in this method and
thereby requests for a special consideration. The entropy conservation
can be taken into account by replacing Eq. (\ref{Vm1}) by the equation
of the entropy conservation, $S = \sigma^* V^*$, where $\sigma^*$ is
the entropy density in droplet-rest frame. This way the volume  $V^*$
becomes an independent variable to be determined from this set of
equations rather than being rigidly defined by the Lorentz
contraction factor $\gamma^*$. It was found out that spectra
calculated with this 
additional requirement of the  entropy conservation coincide with
those based on Eqs. (\ref{Nm1})--(\ref{Pim1})  within 1\%. 
It implies that the entropy is fairly
good conserved already within the modified Milekhin's method defined
by Eqs. (\ref{Nm1})--(\ref{m-FO}).

\begin{figure}[thb]
\includegraphics[width=6.3cm]{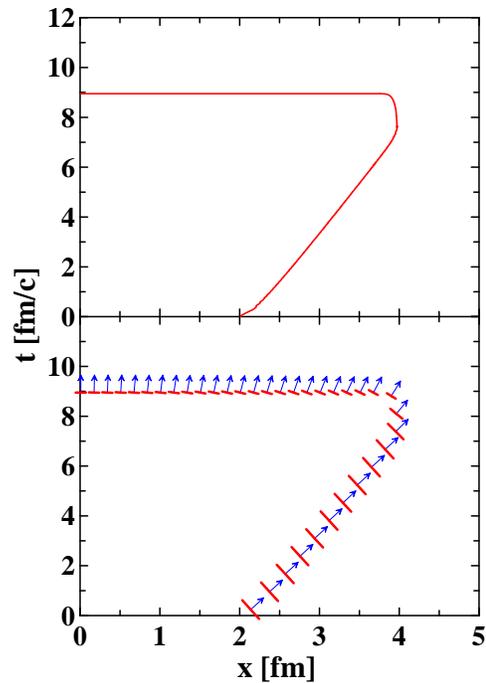}
\caption{(Color online)
Freeze-out hypersurface for hydrodynamic 
evolution the 1D
step-like slab of nuclear matter. 
The upper panel displays the Cooper--Frye choice for the
hypersurface. The lower panel schematically illustrates the modified
Milekhin's prescription, cf. Eq. (\ref{m-FO}), for the
hypersurface. Arrows indicate local 4-velocities on 
this hypersurface. This figure is borrowed from Ref. \cite{3FD-FO}. 
}
\label{fig0}
\end{figure}

It is important that two above methods of subsects. \ref{Computation} and
\ref{Local-Rest} imply that the global freeze-out
hypersurface is in general discontinuous. This hypersurface is
composed of 3-dimensional pieces $\Delta\sigma$ associated with 
weight $(\Delta\sigma \ n_\mu p^\mu)$, with which this droplet is
represented in the total sum over all frozen-out droplets. Here
$n_\mu$ is 
the normal 4-vector to the piece $\Delta\sigma$  of the hypersurface. 
In particular, this weight is 
$V p^0$ in Eq. (\ref{p0-FO}) [$n_\mu=(1,0,0,0)$ in the
computation frame] or $V^* (p_\mu u^{*\mu})$ in
Eq. (\ref{m-FO}) [$n_\mu=u^*_{\mu}$].  An example of such discontinuous
hypersurface in (1+1) dimensions is presented in Fig. \ref{fig0}
(lower panel). 

\subsection{Cooper--Frye Freeze-out}
\label{Cooper--Frye}

The Cooper--Frye hypersurface \cite{Cooper} is constructed on the 
condition that this hypersurface is continuous, see  Fig. \ref{fig0}
(upper panel). In the Cooper--Frye approach parameters of the
distribution function, $\mu$, $T$ and $u^\mu$, are determined from 
Eqs. (\ref{N})--(\ref{Pi}). 
The invariant
spectrum of observable particles is expressed as follows 
\begin{eqnarray}
\label{CF-FO}
\left(E \frac{d N}{d^3 p}\right)_{\scr{Cooper--Frye}} =
\Delta\sigma \ n_\mu p^\mu \ f (p,x),  
\end{eqnarray}
where $n_\mu$ is 
the normal 4-vector to the $\Delta\sigma$ pieces of the continuous
hypersurface. This formula cannot be already associated only with
choice of a reference frame. It can be done, if $n_\mu n^\mu = 1$,
i.e. if  $n^\mu$ is time-like. However, no frame corresponds to 
$n_\mu n^\mu = -1$. Parts of the hypersurface with space-like 
$n^\mu$ are unavoidable consequence of continuity of it. Precisely
with these parts connected is a problem of the  Cooper--Frye
method. If $n_\mu p^\mu < 0$, occurring at space-like 
$n^\mu$, the spectrum of Eq. (\ref{CF-FO}) is negative 
\cite{Sinyukov89,Bugaev96}. This is a severe problem of the method. 
Note that above discussed recipes (\ref{p0-FO}) and (\ref{m-FO}) 
do not reveal this problem.  
\\[5mm]

An important option of the above constructions is weather  
the frozen-out matter is removed from the
hydrodynamic evolution or not. This removal is associated with certain
drain terms, $Q$ and $R^\nu$, in
the r.h.s. of hydrodynamic equations 
\begin{eqnarray}
\label{fl-eq-J}
\partial_\mu J^\mu = Q, 
\\
\label{fl-eq-T}
\partial_\mu T^{\mu\nu} = R^\nu,  
\end{eqnarray}
where $J^\mu$ and $T^{\mu\nu}$ are the baryon current and
energy-momentum tensor, respectively. An example of such drain
terms is presented in Ref. \cite{3FD-FO}.

The Cooper--Frye method unambiguously implies that 
the freeze-out does not affect the hydrodynamic evolution of the
system, i.e. the frozen-out matter is not removed from 
the hydrodynamic phase: $Q_{\scr{CF}}=0$ and $R^\nu_{\scr{CF}}=0$.
The Cooper--Frye freeze-out, which is applied in the
major part of hydrodynamic calculations now, proceeds in the following
way. The hydro calculation runs absolutely unrestricted. The freeze-out
hypersurface is determined by analyzing the resulting 4-dimensional
field of hydrodynamic quantities on the condition of the freeze-out
criterion being met.

\begin{figure}[thb]
\includegraphics[width=7.9cm]{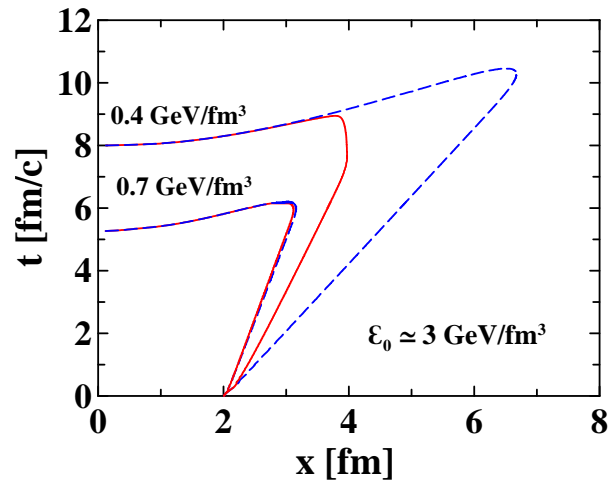}
\caption{(Color online)
Characteristic curves, corresponding to constant values of the energy
density  $\varepsilon$, for hydrodynamic 
evolution the 1D step-like slab of the 4 fm width.
Initial conditions for this slab are constructed on the assumption
that they are formed by the shock-wave mechanism
in head-on collisions of two 1D slabs at
$E_{\scr{lab}}=$ 10 $A$ GeV. Thus constructed initial state
corresponds to the initial energy density $\varepsilon_0\simeq$ 3
GeV/fm$^3$. 
Characteristic curves correspond to 
$\varepsilon= 0.4$ and 0.7 GeV/fm$^3$, 
calculated with (solid lines) 
and without (dashed lines) removal of 
the frozen-out matter from the hydrodynamic evolution. 
This figure is borrowed from Ref. \cite{3FD-FO}. 
}
\label{fig1d}
\end{figure}

At the same time, the modified Milekhin's
method (\ref{m-FO}) and the freeze-out in the
computation frame (\ref{p0-FO}) can be used in both regimes. 
In both cases the energy and momentum are conserved. 
Examples of the modified Milekhin's method with and without removal of 
the frozen-out matter from the hydrodynamic evolution are presented in
Ref.  \cite{3FD-FO}. 
The removal of the matter indeed affects the system evolution. 
This influence is illustrated in Fig. \ref{fig1d}. 
The freeze-out criterion used in this calculation stated that the
matter is frozen-out when the local energy density $\varepsilon$ gets
lower than 0.4 GeV/fm$^3$. The 
$\varepsilon= 0.4$ GeV/fm$^3$ characteristic curves 
calculated with and without freeze-out turn out to be different. 
Note that the value $\varepsilon= 0.4$ GeV/fm$^3$ is achieved right at
the surface of the system, if the frozen-out matter is removed. At
the same time  the $\varepsilon= 0.7$  GeV/fm$^3$ characteristic
curves, which lie quite deep inside the system, remain fairly
unaffected by the freeze-out. 

\section{Discussion}
\label{Discussion}

We considered a puzzle which was in fact a prime motivation 
of the Cooper--Frye \cite{Cooper} approach to the freeze-out in 
relativistic hydrodynamics. 
The puzzle consists in the fact that naive calculation of the total
energy-momentum of unbound equilibrium system does not produce a
4-vector and, moreover, depends on the reference frame. We argue that
a finite unbound system which is equilibrium in one reference frame is in
general nonequilibrium in another frame. This is a consequence of the
relative character of the time synchronization in the relativistic
physics. Thus, naive assumption that this system is equilibrium in any
reference frame results in this puzzle. 
Solution of the puzzle reveals that the
Cooper--Frye recipe is far not a unique phenomenological method that
meets requirements of energy-momentum conservation. Alternative freeze-out
recipes are considered and discussed.

The above discussion concerned precisely {\em phenomenological
 methods}. Recently microscopic treatments of the freeze-out process 
were advanced based on the Bolzmann equation \cite{Sinyukov02,Sinyukov08}
and Kadanoff--Baym equations \cite{Knoll08}. It was found that these 
microscopic approaches approximately justify Cooper--Frye formula
(\ref{CF-FO}) but only on the space-like part of the freeze-out
hypersurface (i.e. possessing a time-like normal vector). Note that on
this part of the hypersurface the Cooper--Frye method is very close to
the modified Milekhin's method (\ref{m-FO}) (cf. Fig. \ref{fig0}) as
well as to the freeze-out in the computation frame (\ref{p0-FO}). 
The Cooper--Frye formula on the time-like part of the freeze-out
hypersurface is not reproduced by these treatments. Precisely on this
part the Cooper--Frye formula essentially differs from two above
mentioned alternative methods and also meets the problem of the
negative spectrum.

Two main conclusions have been drawn from these microscopic
considerations. First, the frozen-out matter 
should be removed from the hydrodynamic evolution. This removal is
important for the total energy-momentum conservation. This conclusion
testifies certainly not in favor of the standard Cooper--Frye method. 
Another basic conclusion is that sharp freeze-out at some 3D
hypersurface is a rather rough approximation to the spectrum
formation, because the freeze-out process is fairly extended in space
and time. It means that the particle emission takes place 
from an extended 4-volume rather than from a 3-dimensional hyposurface as
it is assumed in all above considered phenomenological methods. 
This conclusion is also supported by kinetic simulations, see
e.g. \cite{Pratt08}. Therefore, it makes
all above phenomenological methods questionable. However, the numeric
implementation of the microscopic methods developed in
Refs. \cite{Sinyukov02,Sinyukov08,Knoll08} in 3D hydrodynamic
simulations is highly complicated, because it requires integration
over future evolution of the system for the calculation of the
particle emission at fixed time instant. The implementation performed in
Refs. \cite{Sinyukov08} is not quite consistent, since it does not
take into account the removal of the freeze-out with the hydrodynamic
evolution. Therefore, we still have to use phenomenological methods of
freeze-out in actual hydrodynamic simulations of heavy-ion
collisions. The pending problem is to find out which of the
phenomenological methods most closely simulates results of the 
microscopic methods.

\section*{Acknowledgements}


We are grateful to  S.V. Akkelin, J. Knoll, E.E. Kolomeitsev, 
Yu.M. Sinyukov, V.V. Skokov, V.D. Toneev,  and D.N. Voskresensky for fruitful
discussions.
This work was supported 
the Deutsche
Forschungsgemeinschaft (DFG project 436 RUS 113/558/0-3), the
Russian Foundation for Basic Research (RFBR grant 06-02-04001 NNIO\_a),
Russian Federal Agency for Science and Innovations
(grant NSh-3004.2008.2).

\end{document}